\documentclass[a4paper,11pt]{article}
\usepackage{pos}

\newcommand{\ud}{\mathrm{d}}

\title{3D structure of hadrons and energy-momentum tensor}

\author*[a]{C\'edric Lorc\'e}

\affiliation[a]{CPHT, CNRS, \'Ecole polytechnique, Institut Polytechnique de Paris, \\
91120 Palaiseau, France}

\emailAdd{cedric.lorce@polytechnique.edu}

\abstract{The three-dimensional spatial structure of hadrons is encoded in their form factors. Via appropriate Fourier transform, the latter describe how charge, energy, linear and angular momentum, but also pressure are distributed inside these systems. Electromagnetic form factors of the nucleon have been measured for a long time, but it is only recently that some gravitational form factors have been extracted from experimental data, creating a great deal of enthusiasm in the hadronic community. Here we summarize some of the recent developments on the interpretation of these form factors, and provide a quick overview of what we can learn about the nucleon mass, spin and internal pressure.}

\FullConference{31st International Workshop on Deep Inelastic Scattering (DIS2024)\\
 8–12 April 2024\\
Grenoble, France\\}

\begin{document}
\maketitle

\section{Introduction}

Understanding the inner workings of hadrons requires multidimensional mapping of the distribution of their constituents~\cite{Belitsky:2003nz,Meissner:2009ww,Lorce:2011dv}. Feynman's parton picture has now been promoted to a 5D representation based on the notion of light-front Wigner (or phase-space) distribution~\cite{Lorce:2011kd}, which gives the quasi-probability to find a quark or a gluon with a given light-front three-momentum $(\vec k_\perp,xP^+)$ at the transverse position $\vec b_\perp$. Integrating out some of these variables, one obtains various lower-dimensional distributions which can be probed via different (semi-)inclusive and exclusive processes~\cite{AbdulKhalek:2021gbh}. 

In this contribution, we will focus on the spatial structure as probed by exclusive scattering experiments. The response of hadrons to electromagnetic probes has been measured with extreme precision over the past decades, allowing us to gain insight not only into the electromagnetic structure~\cite{Punjabi:2015bba,Gao:2021sml}, but also into the mechanical structure~\cite{Burkert:2023wzr} of hadrons. More precisely, elastic lepton-hadron scattering gives direct access to the electromagnetic current $\langle p',s'|j^\mu|p,s\rangle$, where $p$ is the four-momentum and $s$ is the polarization of the hadron, whereas deeply virtual Compton scattering provides indirect information on the energy-momentum tensor $\langle p',s'|T^{\mu\nu}|p,s\rangle$. 

Thanks to Poincar\'e symmetry these matrix elements can be described by a limited number of Lorentz-invariant functions of the four-momentum transfer $\Delta=p'-p$ called form factors (FFs). Their Fourier transform in the Breit frame (BF) are usually interpreted as static 3D spatial distributions~\cite{Ernst:1960zza,Sachs:1962zzc,Polyakov:2002yz}. Since relativistic recoil corrections hinder their interpretation as probabilistic densities~\cite{Yennie:1957skg,Burkardt:2000za,Jaffe:2020ebz}, alternative 2D distributions on the light front (LF) have been proposed~\cite{Burkardt:2002hr,Miller:2010nz,Miller:2018ybm}. While the latter admit a strict probabilistic interpretation, they show features that are at odds with the expected picture of hadrons at rest~\cite{Miller:2007uy,Carlson:2007xd}.

To clarify the situation, the concept of relativistic spatial distribution has recently been revisited by a number of authors, see e.g.~\cite{Panteleeva:2021iip,Kim:2021kum,Freese:2021mzg,Epelbaum:2022fjc,Panteleeva:2022khw,Li:2022hyf,Freese:2022fat,Freese:2023abr}. The phase-space approach~\cite{Lorce:2018zpf,Lorce:2018egm,Lorce:2020onh} differs from these attempts by relaxing the requirement of strict probabilistic interpretation to a quasi-probabilistic one. In doing so, it is possible to define relativistic spatial distributions which depend on the hadron average momentum. In this picture, BF and LF distributions appear as two particular limits, and the LF distortions are understood as purely kinematical artifacts associated with spin.

\section{Phase-space formalism}

The LF formalism is closely related to the standard description in the infinite-momentum frame (IMF)~\cite{Kogut:1969xa}, i.e. the frame where the target travels at nearly the speed of light. From a phase-space perspective, elastic frame (EF) distributions~\cite{Lorce:2017wkb,Lorce:2018zpf,Lorce:2018egm} provide a natural interpolation between slow- and fast-moving targets.

The expectation value of an operator $O$ in a physical state $|\psi\rangle$ can be written as
\begin{equation}
    \langle \psi|O|\psi\rangle=\int\frac{\ud^3P}{(2\pi)^3}\,\ud^3R\,\rho_\psi(\vec R,\vec P)\,\langle O\rangle_{\vec R,\vec P},
\end{equation}
where 
\begin{equation}
    \rho_\psi(\vec R,\vec P)=\int\frac{\ud^3q}{(2\pi)^3}\,e^{-i\vec q\cdot\vec R}\,\tilde\psi^*(\vec P+\tfrac{\vec q}{2})\tilde\psi(\vec P-\tfrac{\vec q}{2})
\end{equation}
is the Wigner distribution~\cite{Wigner:1932eb} and $\tilde\psi(\vec p)=\langle p|\psi\rangle/2\sqrt{p^0}$ is the wave packet of the system. The amplitude
\begin{equation}
    \langle O\rangle_{\vec R,\vec P}=\int\frac{\ud^3\Delta}{(2\pi)^3}\,e^{i\vec\Delta\cdot\vec R}\,\frac{\langle p'|O|p\rangle}{2\sqrt{p^0p'^0}}
\end{equation}
can then be interpreted as the expectation value of $O$ for a system localized in phase space with average momentum $\vec P=(\vec p'+\vec p)/2$ and average position $\vec R$. 

To obtain a static distribution in the case of a spacetime-dependent operator $O(x)$, we need to pick up a frame where the energy transfer to the system $\Delta^0=\vec P\cdot\vec \Delta/P^0$ vanishes. This is satisfied, e.g., in the BF defined by the condition $|\vec P|=0$. In the case of a moving target, we can choose for convenience the $z$-axis along $\vec P$. The elastic condition $\Delta^0=0$ is then satisfied if we project the distribution onto the transverse plane
\begin{equation}
    O_\text{EF}(\vec b_\perp;P_z)=\int\ud z\,\langle O\rangle_{\vec R,P_z\vec e_z}(x)=\int\frac{\ud^2\Delta_\perp}{(2\pi)^2}\,e^{-i\vec\Delta_\perp\cdot\vec b_\perp}\,\frac{\langle p'|O(0)|p\rangle}{2P^0}\bigg|_{\Delta_z=|\vec P_\perp|=0}.
\end{equation}
Because of translation symmetry, this EF distribution depends only on the relative transverse position $\vec b_\perp=\vec x_\perp-\vec R_\perp$ besides $P_z$.

For a relativistic system, the $P_z$-dependence cannot in general be factored out. We however expect that~\cite{Jacob:1959at,Durand:1962zza}
\begin{equation}
	\begin{aligned}\label{EFLorentzTrans-Spinj}
		&\langle p',s'|O^{\mu_1\cdots\mu_n}(0)|p,s\rangle = \\
  &\qquad\sum_{s_{B}',s_{B}} {D}^{*(j)}_{s_{B}'s'}(p_{B}',\Lambda){D}^{(j)}_{s_{B}s}(p_{B},\Lambda) \,\Lambda^{\mu_1}_{\phantom{\mu_1}\nu_1}\cdots \Lambda^{\mu_n}_{\phantom{\mu_n}\nu_n}\,\langle p_{B}',s_{B}'|O^{\nu_1\cdots\nu_n}(0)|p_{B},s_{B}\rangle,
	\end{aligned}
\end{equation}
where $\langle p_{B}',s_{B}'|O^{\nu_1\cdots\nu_n}(0)|p_{B},s_{B}\rangle$ is the BF matrix element and $\Lambda^{\mu}_{\phantom{\mu}\nu}$ is the matrix that implements the Lorentz boost from the BF to a generic frame. For a target with spin $j$, the Wigner spin rotation matrix ${D}^{(j)}_{s_Bs}(p_B,\Lambda)$ plays a key role for understanding the aforementioned strange features seen in the LF spatial distributions in terms of distortions induced by the boost~\cite{Lorce:2020onh,Lorce:2022jyi,Chen:2022smg,Chen:2023dxp}. We will restrict the discussion to the case of spin-$1/2$ targets in the following.

\section{Electromagnetic current}

The 3D BF distributions of electric charge and current are defined in the phase-space formalism as
\begin{equation}
    J^\mu_B(\vec r)=\langle j^\mu\rangle_{\vec R,\vec 0}(\vec x)=\int\frac{\ud^3\Delta}{(2\pi)^3}\,e^{-i\vec\Delta\cdot\vec r}\,\frac{\langle p'_B,s'_B|j^\mu(0)|p_B,s_B\rangle}{2P^0_B}
\end{equation}
with $p^\mu_B=(P^0_B,-\vec\Delta/2)$ and $p'^\mu_B=(P^0_B,\vec\Delta/2)$ the initial and final BF four-momenta, and $\vec r=\vec x-\vec R$ the position relative to the center of the system. They differ from the conventional Sachs distributions~\cite{Ernst:1960zza,Sachs:1962zzc} by the normalization factor $2P^0_B$ instead of $2M$ in the denominator. In a similar way, the corresponding 2D EF distributions are given by
\begin{equation}
    J^\mu_\text{EF}(\vec b_\perp;P_z)=\int\ud z\,\langle j^\mu\rangle_{\vec R,P_z\vec e_z}(\vec x)=\int\frac{\ud^2\Delta_\perp}{(2\pi)^2}\,e^{-i\vec\Delta_\perp\cdot\vec b_\perp}\,\frac{\langle p',s'|j^\mu(0)|p,s\rangle}{2P^0}\bigg|_{\Delta_z=|\vec P_\perp|=0}.
\end{equation}
They reduce in the limit $P_z\to 0$ to the projection of the BF distributions onto the transverse plane $J^\mu_\text{EF}(\vec b_\perp;0)=\int\ud z\,J^\mu_B(\vec r)$. As expected, the total electric charge $\mathcal Q=\int\ud^2b_\perp\, J^0_\text{EF}(\vec b_\perp;P_z)\big|_{s'=s}=\langle p,s|j^0(0)|p,s\rangle/(2p^0)$ is independent of $P_z$, which confirms that the momentum-space amplitudes are correctly normalized.

The matrix elements of the electromagnetic current are traditionally parametrized in terms of the so-called Dirac and Pauli FFs
\begin{equation}\label{genparam}
	\langle p',s'|j^\mu(0)|p,s\rangle=e\,\overline u(p',s')\left[\gamma^\mu\,F_1(Q^2)+\frac{i\sigma^{\mu\nu}\Delta_\nu}{2M}\,F_2(Q^2)\right]u(p,s),
\end{equation}
where $Q^2=-\Delta^2$. One of the appealing features of the BF is that the spin structure of the matrix elements is the same as in the non-relativistic theory~\cite{Yennie:1957skg,Ernst:1960zza,Sachs:1962zzc}. One finds that
\begin{equation}\label{BFampl}
	\begin{aligned}
		\langle p'_B,s'_B|j^0(0)|p_B,s_B\rangle&=e\,2M\,\delta_{s'_Bs_B}\,G_E(Q^2),\\
		\langle p'_B,s'_B|\vec j(0)|p_B,s_B\rangle&=e\,(\vec\sigma_{s'_Bs_B}\times i\vec\Delta)\,G_M(Q^2),
	\end{aligned}
\end{equation}
where $\vec\sigma$ are the three Pauli matrices. The combination $G_E=F_1-\tau F_2$ with $\tau=Q^2/(4M^2)$ is called the electric Sachs FF, and $G_M=F_1+F_2$ is called the magnetic Sachs FF. 

The spin structure is more complicated in the EF~\cite{Lorce:2020onh,Kim:2021kum,Chen:2022smg} 
\begin{equation}\label{spinhalfexplicitLT}
    \begin{aligned}
        \langle p',s'|j^0(0)|p,s\rangle\big|_\text{EF}&=e\,2M\,\gamma\,\Bigg[\left(\delta_{s's}\,\cos\theta+\frac{(\vec\sigma_{s's}\times i\vec \Delta_\perp)_z}{|\vec\Delta_\perp|}\,\sin\theta\right)G_E(Q^2)\\
        &\qquad\qquad\quad +\beta\left(-\delta_{s's}\,\sin\theta+\frac{(\vec\sigma_{s's}\times i\vec \Delta_\perp)_z}{|\vec\Delta_\perp|}\,\cos\theta\right)\sqrt{\tau}\,G_M(Q^2)\Bigg],\\
         \langle p',s'|j^3(0)|p,s\rangle\big|_\text{EF}&=e\,2M\,\gamma\,\Bigg[\beta\left(\delta_{s's}\,\cos\theta+\frac{(\vec\sigma_{s's}\times i\vec \Delta_\perp)_z}{|\vec\Delta_\perp|}\,\sin\theta\right)G_E(Q^2)\\
        &\qquad\qquad\quad +\left(-\delta_{s's}\,\sin\theta+\frac{(\vec\sigma_{s's}\times i\vec \Delta_\perp)_z}{|\vec\Delta_\perp|}\,\cos\theta\right)\sqrt{\tau}\,G_M(Q^2)\Bigg],\\
        \langle p',s'|\vec j_\perp(0)|p,s\rangle\big|_\text{EF}&=e\,(\sigma_z)_{s's}\,(\vec e_z\times i\vec \Delta_\perp)\,G_M(Q^2),
    \end{aligned}
\end{equation}
as a result of the Wigner rotation in~\eqref{EFLorentzTrans-Spinj}. The Lorentz boost parameters are given by $\gamma=P^0/P^0_B$ and $\beta=P_z/P^0$, and the Wigner rotation angle $\theta$ satisfies
\begin{equation}\label{WignerAngleSinCos}
    \cos\theta=\frac{P^0+M(1+\tau)}{(P^0+M)\sqrt{1+\tau}},\qquad \sin\theta=-\frac{\sqrt{\tau}P_z}{(P^0+M)\sqrt{1+\tau}}.
\end{equation}
In the limit $P_z\to 0$ one recovers~\eqref{BFampl} with the restriction $\Delta_z=0$, while in the IMF limit $P_z\to\infty$ one finds using the LF components $a^\pm=(a^0\pm a^3)/\sqrt{2}$ 
\begin{equation}\label{LFspinhalfexplicitPM}
    \begin{aligned}
        \langle p',s'|j^+(0)|p,s\rangle\big|_\text{IMF}&=e\,2P^+\,\bigg[\delta_{s's}\,F_1(Q^2)+\frac{(\vec\sigma_{s's}\times i\vec\Delta)_z}{2M}\,F_2(Q^2)\bigg],\\
         \langle p',s'|j^-(0)|p,s\rangle\big|_\text{IMF}&=e\,2P^-\,\bigg[\delta_{s's}\,G_1(Q^2)+\frac{(\vec\sigma_{s's}\times i\vec\Delta)_z}{2M}\,G_2(Q^2)\bigg],\\
        \langle p',s'|\vec j_\perp(0)|p,s\rangle\big|_\text{IMF}&=e\,(\sigma_z)_{s's}\,(\vec e_z\times i\vec \Delta_\perp)\,G_M(Q^2),
    \end{aligned}
\end{equation}
where $G_1=(G_E-\tau G_M)/(1+\tau)$ and $G_2=-(G_E+G_M)/(1+\tau)$. After Fourier transform, these coincide with the 2D LF distributions~\cite{Burkardt:2002hr,Miller:2007uy,Carlson:2007xd}. 

\begin{figure}[t]
\includegraphics[width=\hsize]{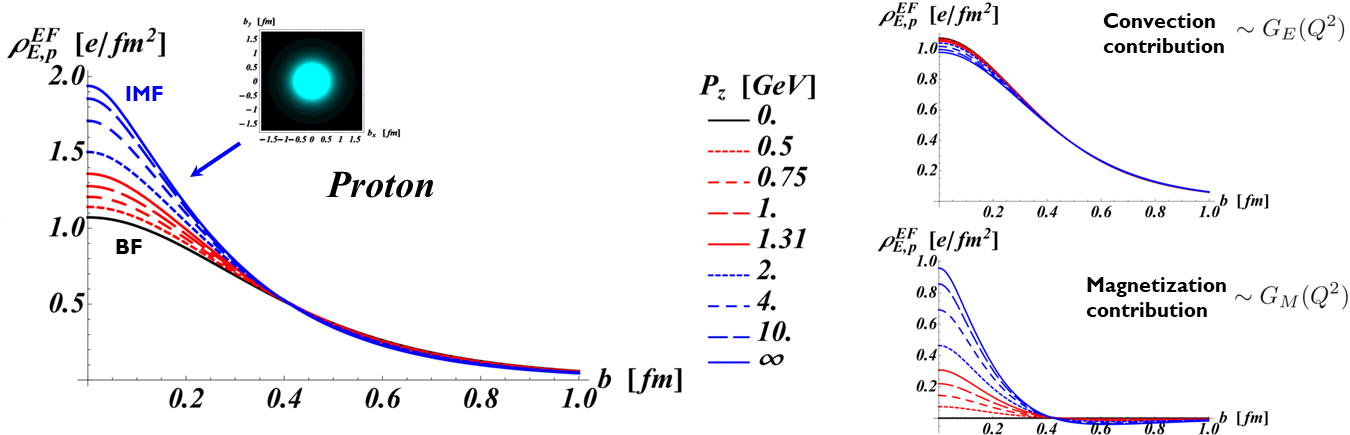}
\includegraphics[width=\hsize]{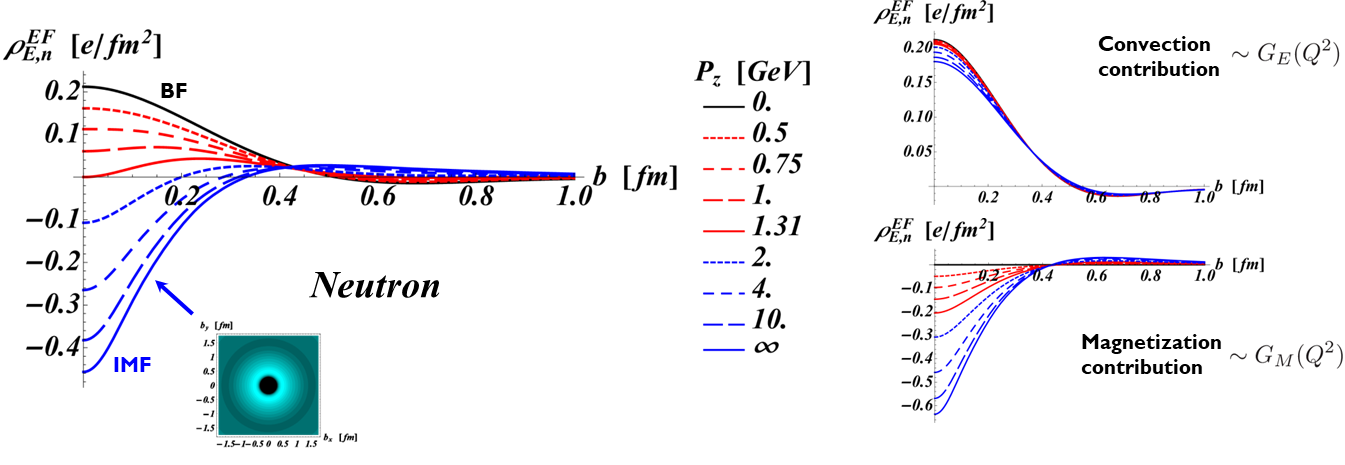}
\caption{Spin-averaged EF charge distributions $\rho^{EF}_E\equiv J^0_\text{EF}(\vec b_\perp;P_z)$ for selected values of the nucleon average momentum. The small panels on the right correspond to the electric (or convective) and magnetic (or magnetization) contributions. Figure adapted from~\cite{Lorce:2020onh}.}\label{fig:rhoE}
\end{figure}

As a result of the combination of a four-vector transformation with a Wigner spin rotation, the unpolarized charge distribution receives a magnetic contribution when $\vec P\neq \vec 0$. Since the induced magnetic contribution in the neutron is large and opposite to the electric contribution, a negatively charged region appears at the center of the LF charge distribution of the neutron, see Fig.~\ref{fig:rhoE}. This is purely an effect of perspective which shows that, although probabilistic, the LF distributions cannot be interpreted as intrinsic densities. 

\section{Energy-momentum tensor}

The energy-momentum tensor (EMT) is a fundamental operator in quantum field theory, for it measures key properties of physical systems like, e.g., mass, spin and pressure. A direct access to its matrix elements requires gravitational scattering, which is far too small to be measured in practice. Some exclusive processes like deeply virtual Compton scattering are described in terms of non-local correlators, known as generalized parton distributions, which can be related to some of the gravitational FFs that parametrize the EMT matrix elements~\cite{Ji:1996ek}. Gravitational FFs can therefore in principle be constrained from experimental data. Applying the same formalism as for the matrix elements of the electromagnetic current, one can define EMT distributions in the BF~\cite{Polyakov:2002yz} and generalize them to both the EF and the LF~\cite{Lorce:2018egm}.

The nucleon matrix elements of the quark EMT can conveniently be parametrized as
\begin{equation}
\begin{aligned}
     \langle p',s'|T^{\mu\nu}_q(0)|p,s\rangle&=\overline u(p',s')\left[\frac{P^\mu P^\nu}{M}\,A_q(Q^2)+\frac{\Delta^\mu\Delta^\nu-g^{\mu\nu}\Delta^2}{4M}\,D_q(Q^2)+Mg^{\mu\nu}\bar C_q(Q^2)\right.\\
     &\qquad\qquad\quad+\left.\frac{P^{\{\mu}i\sigma^{\nu\}\lambda}\Delta_\lambda}{M}\,J_q(Q^2)-\frac{P^{[\mu}i\sigma^{\nu]\lambda}\Delta_\lambda}{M}\,S_q(Q^2)\right]u(p,s),
\end{aligned}
\end{equation}
where the first four terms describe the symmetric part $T^{\{\mu\nu\}}=(T^{\mu\nu}+T^{\nu\mu})/2$ while the fifth term accounts for the antisymmetric part $T^{[\mu\nu]}=(T^{\mu\nu}-T^{\nu\mu})/2$. A similar parametrization holds for the gluon EMT. Poincar\'e symmetry implies conservation of linear and angular momentum, as well as mechanical equilibrium, which amount to the following constraints on the gravitational FFs
\begin{equation}
        A_q(0)+A_g(0)=1,\qquad
        J_q(0)+J_g(0)=\tfrac{1}{2},\qquad \bar C_q(Q^2)+\bar C_g(Q^2)=0.
\end{equation}

In the BF, one finds that
\begin{equation}\label{BFEMTampl}
	\begin{aligned}
		\langle p'_B,s'_B|T^{00}_q(0)|p_B,s_B\rangle&=2MP^0_B\,\delta_{s'_Bs_B}\left\{A_q(Q^2)+\bar C_q(Q^2)+\tau[D_q(Q^2)-B_q(Q^2)]\right\},\\
		\langle p'_B,s'_B|T^{\{0k\}}_q(0)|p_B,s_B\rangle&=2P^0_B\,(\vec\sigma_{s'_Bs_B}\times i\vec\Delta)^k\,J_q(Q^2),\\
  \langle p'_B,s'_B|T^{[0k]}_q(0)|p_B,s_B\rangle&=-2P^0_B\,(\vec\sigma_{s'_Bs_B}\times i\vec\Delta)^k\,S_q(Q^2),\\
  \langle p'_B,s'_B|T^{ij}_q(0)|p_B,s_B\rangle&=2MP^0_B\,\delta_{s'_Bs_B}\left\{\frac{\Delta^i\Delta^j}{4M^2}\,D_q(Q^2)-\delta^{ij}[\bar C_q(Q^2)+\tau D_q(Q^2)]\right\},
	\end{aligned}
\end{equation}
where $B_q=2J_q-A_q$. Determining these matrix elements constitute one of the main motivations of the physics program of the future Electron-Ion Collider~\cite{AbdulKhalek:2021gbh}. Here are some examples of the fundamental physical properties they encode:
\begin{itemize}
    \item The amplitude $\langle p'_B,s'_B|T^{00}_q(0)|p_B,s_B\rangle/(2P^0_B)$ is related via Fourier transform to the spatial distribution of the quark energy $T^{00}_{q,B}(\vec r)$. In the forward limit $\Delta\to 0$, it reduces to the quark contribution to the nucleon mass~\cite{Lorce:2017xzd,Lorce:2021xku}. 
    \item The spatial distribution of quark Belinfante angular momentum is obtained from
\begin{equation}
    \mathcal J^i_{q,B}(\vec r)=\epsilon^{ijk}r^j T^{\{0k\}}_{q,B}(\vec r),
\end{equation}
while the spatial distribution of quark kinetic angular momentum is given by~\cite{Leader:2013jra,Lorce:2017wkb}
\begin{equation}
    J^i_{q,B}(\vec r)=L^i_{q,B}(\vec r)+S^i_{q,B}(\vec r)
\end{equation}
with the orbital and spin contributions defined as
\begin{equation}
    \begin{aligned}
        L^i_{q,B}(\vec r)=\epsilon^{ijk}r^j T^{0k}_{q,B}(\vec r),\qquad
        S^i_{q,B}(\vec r)=\frac{1}{2}\langle\overline\psi\gamma^i\gamma_5\psi\rangle_{\vec 0,\vec 0}(\vec r). 
    \end{aligned}
\end{equation}
\item Because of spherical symmetry about the center of the nucleon, the quark stress tensor is described by two functions~\cite{Polyakov:2002yz,Polyakov:2018zvc,Lorce:2018egm}
\begin{equation}
    T^{ij}_{q,B}(\vec r)=\delta^{ij}\,p_q(r)+\left(\frac{r^i r^j}{r^2}-\frac{1}{3}\,\delta^{ij}\right)s_q(r).
\end{equation}
Since the BF distributions are static, EMT conservation amounts to $\nabla^iT^{ij}_B(\vec r)=0$ and leads to a relation between the total isotropic pressure $p(r)$ and pressure anisotropy $s(r)$
\begin{equation}
    \frac{\ud}{\ud r}\bigg(p+\frac{2}{3}\,s\bigg)+\frac{2}{r}\,s=0.
\end{equation}
Mechanical equilibrium of the nucleon implies the von Laue condition $\int\ud^3r\,p(r)=0$. Stability arguments further suggest that the quantity $D_q(0)+D_g(0)=M\int\ud^3r\,r^2\,p(r)$ should be negative~\cite{Perevalova:2016dln}. 
\end{itemize}
The first extraction of the pressure distribution inside a proton from experimental data on deeply virtual Compton scattering has been reported in~\cite{Burkert:2018bqq}, and followed by more conservative analyses~\cite{Kumericki:2019ddg,Dutrieux:2021nlz}. Data on $J/\psi$-photoproduction in the threshold region have also been used to extract gluon gravitational FFs~\cite{Duran:2022xag}. Recent calculations of the gravitational FFs from Lattice QCD can be found for example in~\cite{Alexandrou:2020sml,Pefkou:2021fni,Wang:2021vqy}. More details and references can be found in the review~\cite{Burkert:2023wzr}.

\section{Conclusions}

Electromagnetic and gravitational form factors encode in a Lorentz-invariant key information about the 3D structure and physical properties of hadrons. When relativistic recoil effects cannot be neglected, spatial distributions become frame-dependent. The relativistic notion of spatial distribution is then naturally formulated in a phase-space approach, where the strict probabilistic interpretation is lost. From this perspective, it is possible to interpolate between the Breit frame ($P_z=0$) and light-front ($P_z\to\infty$) pictures, and to show that the appearance of spatial distributions in a moving frame is strongly impacted by the spin of the target.

\end{document}